\documentclass[sigconf,nonacm,natbib=true]{acmart}
\AtBeginDocument{%
  }

\usepackage{csquotes} 
\usepackage{pbalance}
\usepackage{mdframed}
\usepackage{enumitem}
\usepackage{tabularx}   

\newmdenv[
  backgroundcolor=gray!8,
  linecolor=gray!50,
  roundcorner=4pt,
  skipabove=6pt,
  skipbelow=6pt,
  innerleftmargin=8pt,
  innerrightmargin=8pt,
  innertopmargin=6pt,
  innerbottommargin=6pt
]{prompt}

\makeatletter
\newcommand{\acmrightssize}{\fontsize{8}{9.5}\selectfont}

\newcommand{\firstpagerights}[1]{%
  \begingroup
    \renewcommand\thefootnote{}%
    \footnotetext{%
      \acmrightssize
      \raggedright
      \setlength{\parskip}{0pt}%
      \setlength{\parindent}{0pt}%
      #1%
    }%
    \addtocounter{footnote}{0}%
  \endgroup
}
\makeatother

\begin{document}

\title[OwlerLite: Scope- and Freshness-Aware Web Retrieval for LLM Assistants]{OwlerLite: Scope- and Freshness-Aware \\ Web Retrieval for LLM Assistants}

\author{Saber Zerhoudi}
\orcid{0000-0003-2259-0462}
\affiliation{%
  \institution{University of Passau}
  \city{Passau}
  \country{Germany}
}
\email{saber.zerhoudi@uni-passau.de}

\author{Michael Dinzinger}
\orcid{0009-0003-1747-5643}
\affiliation{%
  \institution{University of Passau}
  \city{Passau}
  \country{Germany}
}
\email{michael.dinzinger@uni-passau.de}

\author{Michael Granitzer}
\orcid{0000-0003-3566-5507}
\affiliation{%
  \institution{University of Passau}
  \city{Passau}
  \country{Germany}
}
\affiliation{%
  \institution{IT:U Austria}
  \city{Linz}
  \country{Austria}
}
\email{michael.granitzer@uni-passau.de}

\author{Jelena Mitrovi\'{c}}
\orcid{0000-0003-3220-8749}
\affiliation{%
  \institution{University of Passau}
  \city{Passau}
  \country{Germany}
}
\email{jelena.mitrovic@uni-passau.de}

\renewcommand{\shortauthors}{S. Zerhoudi et al.}

\begin{abstract}
Browser-based language models often use retrieval-augmented generation (RAG) but typically rely on fixed, outdated indices that give users no control over which sources are consulted. This can lead to answers that mix trusted and untrusted content or draw on stale information. We present OwlerLite, a browser-based RAG system that makes user-defined scopes and data freshness central to retrieval. Users define reusable scopes---sets of web pages or sources---and select them when querying. A freshness-aware crawler monitors live pages, uses a semantic change detector to identify meaningful updates, and selectively re-indexes changed content. OwlerLite integrates text relevance, scope choice, and recency into a unified retrieval model. Implemented as a browser extension, it represents a step toward more controllable and trustworthy web assistants.
\end{abstract}



\begin{teaserfigure}
  \begin{center}
\includegraphics[width=0.9\textwidth]{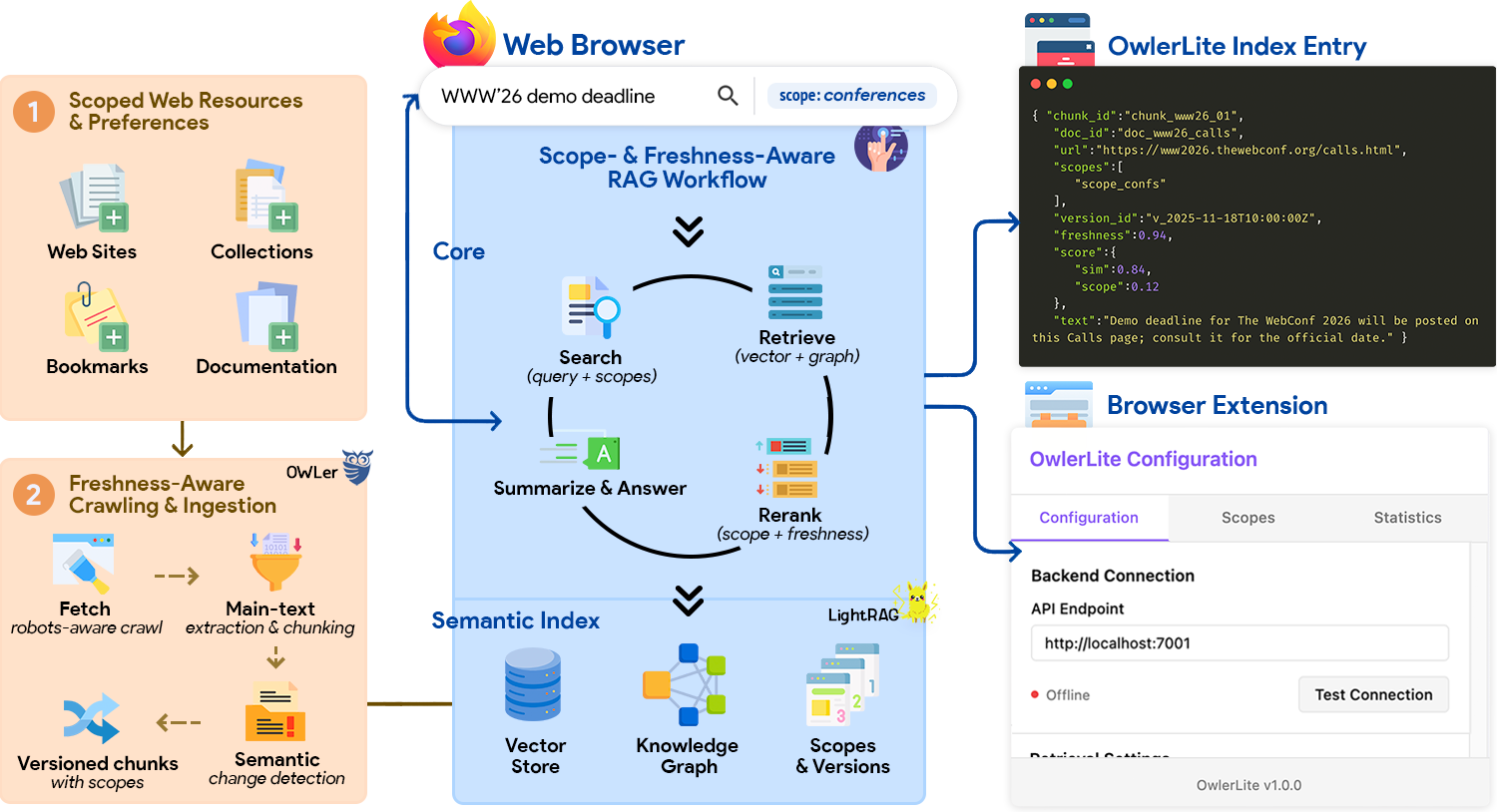}
\vspace{-3mm}
  \caption{High-Level Architecture of OwlerLite with Scope- and Freshness-Aware RAG Workflow.}
  \label{fig_simiir3}
  \vspace{4mm}
\end{center}
\end{teaserfigure}

\maketitle
\enlargethispage{2\baselineskip}
\firstpagerights{%
  © ACM, 2026. This is the author's version of the work.\\
  The definitive version was published in:
  \emph{Proceedings of the Companion Proceedings of the ACM Web Conference 2026 (WWW Companion '26), \\
  April 13--17, 2026, Dubai, United Arab Emirates}.\\
  DOI: \url{https://doi.org/10.1145/3774905.3793140}
}

\section{Introduction}
Browser-integrated LLM assistants aim to improve knowledge by synthesizing information from varied web sources into a single interface~\cite{Nakano:2021:WebGPT,Liu:2023:WebGLM}. Retrieval-augmented generation (RAG) is the standard mechanism for this, where a retriever selects relevant text for an LLM to synthesize an answer~\cite{Lewis:2020:NIPS}. Deploying RAG over live web content, however, introduces two critical challenges.

First, \textbf{scope control} is limited. Many systems use a single large index. Users cannot easily restrict a query to a specific set of trusted resources (e.g, official documentation, bookmarked reading list). This lack of explicit scoping makes it difficult to ensure answer provenance or adhere to organizational policies.

Second, \textbf{freshness} is often treated as a simple operational task. Pages are re-indexed periodically, but this process rarely distinguishes between superficial layout changes and small, semantically important content edits. Consequently, assistants may provide answers from outdated snapshots, even when the live page contains a critical update. Recent work on dynamic and temporally-aware QA highlights how sensitive LLM-based systems are to stale or time-mismatched evidence~\cite{Kasai:2022:RealTimeQA,Gade:2024:TempRALM}.

Existing RAG frameworks (e.g, LightRAG~\cite{Guo:2024:arXiv}) perform well on static corpora but do not treat scope or freshness as explicit modeling features. In parallel, general-purpose LLM assistants with web search capabilities offer broad coverage but remain stateless and opaque. Users cannot define persistent, reusable corpora or control which version of a page is used for an answer.

This paper introduces \textbf{OwlerLite}~\footnote{\label{fn:toolkit} Github Repository: \url{https://github.com/searchsim-org/owlerlite}}, a system that provides a persistent, scope- and freshness-aware RAG layer over curated slices of the Web. Implemented as a browser extension and a backend service using LightRAG, OwlerLite allows users to define named \textit{scopes} corresponding to sets of web resources. A freshness-aware crawler monitors these pages, using text-level similarity to detect only semantically meaningful changes and selectively re-ingest affected text chunks. The retrieval function combines semantic similarity, scope priors, and recency features. Explanations expose which scopes and page versions contributed to each answer.

OwlerLite builds on the OWLer crawler~\cite{Zerhoudi:2023:Zenodo}, developed in the OpenWebSearch.eu project~\cite{Granitzer:2024:JASIS}. While OWLer focuses on large-scale, distributed crawling for a general open web index, OwlerLite provides a complementary, lightweight specialization. It adapts the OWLer philosophy for the browser, focusing on how an assistant can curate, scope, and maintain the freshness of smaller, user-driven corpora for a RAG pipeline.

Our contributions are threefold: (1) we formalize the scope- and freshness-aware RAG problem over web resources, (2) we propose an architecture for semantic change detection and a scope-aware retrieval objective that can be added to LightRAG, and (3) we position OwlerLite relative to other systems and outline an evaluation experiment using MSMARCO~\cite{Nguyen:2016:NIPS,Pradeep:2024:Ragnarok} index with synthetic scopes. 

\section{Related Work}
\paragraph{Retrieval-Augmented Generation and Knowledge Graph RAG}
Retrieval-Augmented Generation (RAG) combines neural retrieval with sequence generation for knowledge-intensive tasks~\cite{Lewis:2020:NIPS}. Standard RAG architectures use a dense retriever over a static corpus. LightRAG~\cite{Guo:2024:arXiv} extends this approach by building a knowledge graph, enabling dual retrieval over both text embeddings and structured graph relationships. Similar knowledge-graph-extended RAG architectures have been proposed for domain-specific QA and multi-hop reasoning~\cite{Xu:2024:KGRAG,Zhu:2025:KG2RAG,Han:2024:GraphRAG}. OwlerLite builds on the LightRAG backend but adapts the retrieval objective and indexing procedures to support user-defined scopes and semantic freshness.

\paragraph{Web Crawling, OWLer, and Freshness}
Web crawling strategies and freshness estimation are well-studied fields, covering incremental crawling and change-rate estimation~\cite{Cho:2003:Refresh,Olston:2010:CrawlingSurvey,Avrachenkov:2020:ChangeRate,Gossen:2015:iCrawl}. The OWLer crawler~\cite{Zerhoudi:2023:Zenodo}, built on Apache Storm, introduced ``scopes of interest'' at the URL frontier. This allows different agents to build topic-specific indices within a large, shared crawl. OwlerLite is inspired by this concept but applies it at a different scale. Instead of partitioning a large-scale crawl, OwlerLite manages per-user scopes over a smaller, curated corpus and integrates this directly into a RAG pipeline. Its semantic change detection is thus a lightweight, RAG-focused complement to large-scale infrastructure like OWLer.

\paragraph{Browser-Based Assistants and Personal Knowledge Systems}
Many tools allow users to index and query local or web content, often providing basic scoping by folder or workspace. These systems typically treat scope and freshness as implementation details rather than formal concepts, despite recent work on personal knowledge graphs and personal information assistance~\cite{Skjaeveland:2024:PKGSurvey,Bakhshizadeh:2024:PKAssist}. OwlerLite's contribution is an explicit formulation of scope fidelity and semantic freshness for these assistants, along with an architecture that connects these notions to a knowledge-graph-based RAG backend.

\paragraph{Web-Search-Enabled LLM Assistants}
Commercial LLM assistants often use live web search to answer queries~\cite{Nakano:2021:WebGPT,Microsoft:2024:Copilot}. These systems are effective for ad-hoc information seeking but operate in a stateless manner. They do not maintain persistent user-defined corpora, nor do they support reusable named scopes for project-specific collections. Their freshness is tied to the search engine, and they do not track semantic versions of content. OwlerLite should be seen as a complementary layer. It maintains a persistent, scope-aware, and version-aware index over a user's \textit{curated} web resources.

\paragraph{Explainable IR and RAG Transparency}
Explainability in information retrieval often involves justifying rankings by highlighting important terms. In RAG systems, this usually takes the form of citations, with recent work proposing dedicated explanation frameworks for both document-centric and graph-based RAG pipelines~\cite{Sudhi:2024:RAGEx,Moghaddam:2025:KGSMILE}. OwlerLite adds scope and version as core explanatory dimensions. Its interface can decompose a retrieval score into its semantic, scope, and freshness components. 

\section{Problem Formulation and Design Principles}
\label{sec:design-principles}
We model a user interacting with an assistant in a web browser. Let $\mathcal{D}$ be the universe of all crawlable web pages.

\textbf{Retrieval scopes.} A scope ($S \subseteq \mathcal{D}$) is a subset of documents defined by the user, often through URL patterns or page selections. Users can define multiple scopes ($S_1, \dots, S_M$) and select a set $\mathcal{S}_q$ at query time. The intended corpus $\mathcal{D}_q$ for a query $q$ is: 
$\textstyle \mathcal{D}_q = \bigcup_{S \in \mathcal{S}_q} S.$

\textbf{Semantic freshness.} The crawler observes each document $d \in \mathcal{D}$ as a sequence of versions ($d^{(0)}, d^{(1)}, ...$) at times $t^{(0)} < t^{(1)} < ...$ . Let $\phi(d^{(i)})$ be the extracted textual content from version $d^{(i)}$. We use a similarity function $\sigma(\cdot,\cdot)$, based on locality-sensitive hashing, to identify a \textit{semantic change} when: $\textstyle 1 - \sigma(\phi(d^{(i)}), \phi(d^{(i+1)})) \> \tau,$ where $\tau$ is a threshold calibrated to ignore superficial edits. A page is \textit{semantically fresh} if its indexed version matches the latest observed version and no semantic change has been detected since indexing.

\textbf{Scope fidelity and scope leakage.} For a query $q$ with corpus $\mathcal{D}_q$ and a ranked list of retrieved passages ($p_1, \dots, p_k$), we define:
\begin{itemize}[label=---,leftmargin=1.7em, itemsep=2pt, topsep=2pt] 
    \item \textit{Scope fidelity at $k$:}
    $\textstyle  \text{SF@}k = \frac{1}{k} \sum_{i=1}^k \mathbf{1}[p_i \in \mathcal{D}_q] $
    \item \textit{Scope leakage at $k$:}
    $\textstyle \text{SL@}k = 1 - \text{SF@}k = \frac{1}{k} \sum_{i=1}^k \mathbf{1}[p_i \notin \mathcal{D}_q]$
\end{itemize}

\textbf{Stale-answer risk.} The \textit{stale-answer risk} $R(q,t)$ is the probability that an answer, generated at time $t$ from a set of indexed passages $P_t$, is outdated. This occurs if the live versions of the source documents contain semantic updates not yet indexed. Our index maintenance strategy aims to reduce this risk by prioritizing re-crawls of pages that show frequent, meaningful updates.

From these definitions, we derive three design principles:
\begin{itemize}[leftmargin=1.7em, itemsep=2pt, topsep=2pt]
    \item \textbf{P1: Scope fidelity.} The system must maximize SF@k and minimize SL@k for scoped queries, subject to relevance.
    \item \textbf{P2: Semantic freshness.} The indexing strategy must focus its crawl budget on pages with a higher stale-answer risk.
    \item \textbf{P3: Transparency.} The system must show users the scope and version information for each retrieved passage.
\end{itemize}

OwlerLite applies these principles via semantic change detection, selective re-ingestion, and a scope- and freshness-aware interface.

\section{System Architecture and Methods}
\subsubsection*{Overview}
OwlerLite comprises three main components: (1) a freshness-aware crawler that monitors web resources within user-defined ``scopes'', detects semantic changes, and selectively re-ingests updated text chunks; (2) a LightRAG-based retrieval backend that stores vector and knowledge graph representations annotated with scope and version metadata; and (3) a browser extension for users to manage scopes and inspect versioned scoped answers.

The system operates on a persistent, user-curated index, unlike assistants that use temporary queries to external search engines. Existing LightRAG deployments can adopt this functionality with minimal changes. Scopes are implemented as metadata filters, and freshness is added as a feature in the scoring function.

\subsubsection*{Freshness-Aware Crawling and Semantic Change Detection}
For each scope, OwlerLite tracks URLs provided by the user. The crawler periodically fetches these URLs, respecting \texttt{robots.txt} and using HTTP validators (i.e., ETag, Last-Modified) where available.

Upon fetching a page, the system performs a three-step ingestion process. First, it extracts the main article text using the Trafilatura extractor~\cite{Barbaresi:2021:Trafilatura}, a DOM-based heuristic method with proven benchmark performance~\cite{Gupta:2022:ContentExtraction}. Second, the cleaned text is segmented into 200--400 token chunks using paragraph and heading boundaries with a small overlap. This size is chosen for its strong recall-efficiency balance in RAG systems~\cite{Chroma:2024:Chunking,Merola:2025:ChunkingRAG}. Third, the system computes a 64-bit SimHash fingerprint $s(c)$ for each chunk $c$ over 5-gram shingles~\cite{Charikar:2002:SimHash,Manku:2007:NearDuplicates}, creating a compact representation for fast, large-scale near-duplicate detection.

To detect updates, the system aligns chunks from new and old page versions and compares their SimHash fingerprints using the Hamming distance. We define a hash-based similarity:
$$
\sigma(c_\text{old}, c_\text{new}) = 1 - \frac{\operatorname{Hamming}\!\bigl(s(c_\text{old}), s(c_\text{new})\bigr)}{64}.
$$
Chunks with $\sigma$ above a high threshold $\tau_1$ (e.g., $0.97$) are considered unchanged. Chunks with $\sigma$ below a lower threshold $\tau_2$ (e.g., $0.90$) are treated as updated. Chunks falling within the intermediate band $[\tau_2, \tau_1]$ undergo a second, embedding-based check. This check uses a semantic deduplication scheme, similar to SemDeDup~\cite{Abbas:2023:SemDeDup}, to identify semantic near-duplicates that differ lexically. Only chunks classified as changed by this two-stage procedure are re-ingested with a new version identifier. This lexical-plus-semantic method filters out minor, non-semantic modifications and retains meaningful semantic rewrites relevant for retrieval.

\subsubsection*{Indexing with Scope and Version Metadata}
During ingestion, each chunk is annotated in LightRAG with its scope identifiers, source URL, a version identifier, an update timestamp, and derived freshness features. These annotations are stored alongside the vector and knowledge graph records. This allows for filtering and explanation at retrieval time.

\subsubsection*{Scope- and Freshness-Aware Retrieval}
\label{sec:scope-freshness-aware}
Given a query $q$ and selected scopes $\mathcal{S}_q$, retrieval occurs in two stages.
First, \textbf{candidate generation} invokes the LightRAG retriever with a metadata filter, restricting results to chunks belonging to $\mathcal{S}_q$.
Second, \textbf{scoring and ranking} are applied to each candidate passage $p$. The final score $h(q,p)$ is a weighted combination of four components:
\begin{itemize}[label=---,leftmargin=1.7em, itemsep=2pt, topsep=2pt] 
    \item $\text{sim}_{\text{vec}}(q,p)$: Semantic similarity from vector embeddings.
    \item $\text{score}_{\text{graph}}(q,p)$: Graph-based evidence score from LightRAG.
    \item $g(p; \mathcal{S}_q)$: A scope prior, which boosts passages within the selected scopes.
    \item $\text{fresh}(p)$: A recency feature, such as a decayed function of time since the last update.
\end{itemize}
The final score is: 

$\textstyle
h(q,p)=\alpha\cdot\text{sim}_{\mathrm{vec}}(q,p)
+(1-\alpha)\cdot\text{score}_{\mathrm{graph}}(q,p)
+\beta\cdot\log g(p;\mathcal{S}_q)
+\delta\cdot\text{fresh}(p)$

\vspace{0.8em}
The weights $\alpha, \beta, \delta$ control the trade-off between relevance, scope fidelity, and recency, aiming to select recent in-scope passages.

\subsubsection*{Explanation and Interaction}
The browser extension uses the retrieval metadata to construct an explanation view. This includes a breakdown of the four score components, scope labels, and a \textit{version lineage view} that shows a compact ``diff'' of semantic changes between versions. An optional natural-language rationale summarizing the selection reasoning can also be generated.

\subsubsection*{Implementation Notes}
The system is modular. The frontend is a TypeScript browser extension. The backend is a Python FastAPI service that wraps LightRAG, the vector store, and the change detector. The LLM for answer generation is accessed via a pluggable API, supporting local or remote models. This architecture supports the use of a Go-based OWLer crawling variant as a backend source.

\section{Offline Evaluation}
OwlerLite\textsuperscript{\ref{fn:toolkit}} is primarily a system contribution. It is desirable to quantify to what extent its scope-aware scoring improves \textit{scope fidelity} without severely degrading classical relevance metrics. Running bespoke user studies is impractical in early stages. We thus outline a minimal offline evaluation protocol using the \textbf{TREC 2024 RAG} corpus, built from the deduplicated and segmented MS~MARCO V2.1 document collection~\cite{Nguyen:2016:MSMARCO,Pradeep:2024:Ragnarok,TREC2024:RAGCorpus,Thakur:2025:SupportEval} and contains precomputed embeddings using Cohere’s \texttt{\small embed-english-v3.0} model~\cite{Cohere:EmbedV3,Reimers:2024:DiskVectorIndex}.

\subsubsection*{Experimental Setup}
\textbf{Corpus and Index.} We use a local OpenSearch index of the TREC 2024 RAG segmented corpus---i.e., the MS MARCO V2.1 segment collection with URL, title, headings, and segment text fields, along with the associated 1024-dimensional embeddings exposed for each chunk~\cite{TREC2024:RAGGuidelines,TREC2024:RAGCorpus}.

\textbf{Queries.} We sample 200 development topics from the TREC 2024 RAG Retrieval (R) task, using the official topic file and relevance assessments provided by the track organizers~\cite{Pradeep:2024:Ragnarok,TREC2024:RAGGuidelines}. We retain only topics with at least one relevant segment in the judged pool.

\textbf{Synthetic Scopes.} To simulate user-defined scopes, we partition the segment corpus into $K = 20$ synthetic groups. We apply $k$-means clustering on document-level embeddings obtained by averaging the segment embeddings belonging to the same document. This assigns each segment to one of $K$ clusters ($S_1, \dots, S_K$), which we treat as synthetic scopes. For a given query $q$, we identify the set of judged relevant segments. The scope containing the largest number of relevant segments is designated as the \textbf{target scope} $S^*_q$. This defines an intended corpus $\mathcal{D}_q = S^*_q$ for each query, modeling a user's intent to search within a specific subset.

\textbf{Systems.} We compare two retrieval configurations:
\begin{itemize}[label=---,leftmargin=1.7em, itemsep=2pt, topsep=2pt] 
    \item \textit{Baseline:} A dense retriever over the \texttt{\small embed-english-v3.0} embeddings. We embed the query $q$ with the same model and retrieve the top-100 segments according to cosine similarity, matching the TREC RAG top-$k$ setting for the Retrieval task~\cite{TREC2024:RAGGuidelines}.
    \item \textit{Scope-aware:} The same dense retriever, but the top-100 candidates are re-ranked using our scoring function $h(q,p)$ from Section~\ref{sec:scope-freshness-aware}. We apply a scope prior $g$ favoring the target scope $S^*_q$ ($\gamma = 0.1$, $\alpha = 0.8$, $\beta = 0.2$, $\delta = 0$).
\end{itemize}

\textbf{Metrics.} We compute relevance using \text{\small NDCG@10} with the official TREC RAG graded relevance labels~\cite{Jarvelin:2002:CG}. In addition, we compute SF@10 and SL@10 (see Section~\ref{sec:design-principles}) relative to the target scope $\mathcal{D}_q$.

\subsubsection*{Results}
Table~\ref{tab:offline-eval} reports the average metrics over the evaluated queries. The scope-aware system increases scope fidelity (SF@10) from 0.64 to 0.83. It also roughly halves scope leakage (SL@10) from 0.36 to 0.17. This gain in scope adherence is achieved with only a minor decrease in relevance (NDCG@10 from 0.503 to 0.495). This suggests that the scope prior successfully guides the retriever to the intended scope while largely preserving topical relevance. For queries where most relevant items are already in the target scope ($\geq 80\%$), the drop in NDCG@10 is negligible ($< 0.005$).

\vspace{-0.6em}
\begin{table}[h!]
\centering
\caption{Retrieval performance on TREC 2024 RAG topics.}
\label{tab:offline-eval}
\vspace{-0.4em}
\begin{tabular}{@{}lccc@{}}
\toprule
\textbf{System} & \textbf{NDCG@10} & \textbf{SF@10} & \textbf{SL@10} \\
\midrule
Baseline    & 0.503 & 0.64 & 0.36 \\
Scope-aware & 0.495 & 0.83 & 0.17 \\
\bottomrule
\end{tabular}
\vspace{-0.6em}
\end{table}

\subsubsection*{Discussion}
These results indicate that our scope-aware scoring function can meaningfully improve scope fidelity on a large-scale corpus with minimal loss in traditional relevance. The experiment remains limited: the scopes are synthetic and we only use a subset of topics. Nevertheless, it provides initial empirical support for our approach and illustrates that \text{\small SF@k} and \text{\small SL@k} can be informative to report alongside standard ranking metrics such as \text{\small NDCG@k}.

\section{Demonstration}




We aim to present OwlerLite as a browser-based assistant running on a single laptop. A local backend will host the LightRAG instance, data stores, and a local LLM endpoint. This self-contained approach ensures robust operation without external dependencies. The demo will use a pre-indexed content set with predefined scopes.

The demonstration flow will (i) introduce the concepts of scopes and semantic freshness, (ii) issue a query without scoping to show baseline RAG behavior, and (iii) rerun the query with one or more selected scopes. The explanation panel will then show how scope priors and freshness signals contributed to the ranking. We will also (iv) display the version history for a passage to highlight semantic changes. This flow will concretely showcase OwlerLite's main contributions: explicit user control over retrieval scopes, semantic freshness-aware indexing, and an explanation layer that reveals the scope and version of each passage.

\section{Conclusion and Future Directions}
OwlerLite\textsuperscript{\ref{fn:toolkit}} is a browser-based RAG system that models retrieval scopes and semantic freshness for curated web resources. It combines semantic change detection with a scope-aware retrieval objective in LightRAG to reduce stale-answer risk and improve scope fidelity, and it exposes scope and version history to support transparent interaction. OwlerLite is designed to complement, rather than replace, web-search LLM assistants by offering a persistent, scope-aware layer for assistant workflows.

We formalized the problem of scope- and freshness-aware RAG, proposed a system design, and reported a small-scale offline evaluation on the MS~MARCO V2.1 corpus using data from the TREC 2024 RAG Track. OwlerLite adapts ideas from the large-scale OWLer crawler~\cite{Zerhoudi:2023:Zenodo} to the context of a browser-based assistant.

The current prototype focuses on a lightweight, heuristic change-detection pipeline (SimHash over n-grams with fixed thresholds $\tau_1, \tau_2$), simple chunk-level recency features, and synthetic scopes obtained from unsupervised clustering on MS~MARCO segments and a subset of TREC 2024 RAG topics. This design keeps the system practical while leaving room for richer learned models of change, scope, and stale-answer risk, as well as evaluations that more closely match real user-defined scopes and organizational policies. OwlerLite also surfaces scope and version data while intentionally leaving content verification and policy decisions to system operators.

Future work will expand the evaluation, replace parts of the heuristic pipeline with learned or domain-specific models, and explore improved combinations of scope and freshness signals. We also plan to integrate OwlerLite with large-scale crawlers such as OWLer and to study collaborative scope management. We hope OwlerLite provides a foundation for further research on controllable, transparent RAG systems in the browser.

\bibliographystyle{ACM-Reference-Format}
\bibliography{sample-base}

\end{document}